\documentstyle[pre,aps,multicol,epsfig]{revtex}
\input{epsf}
\begin{document}
\draft
\title{Decay of the classical Loschmidt echo 
in integrable systems} 
\author{Giuliano Benenti$^{(1)}$, Giulio Casati$^{(1,2)}$, 
and Gregor Veble$^{(1,3)}$}  
\address{$^{(1)}$Center for Nonlinear and Complex  
Systems, Universit\`a degli Studi dell'Insubria and} 
\address{Istituto Nazionale per la Fisica della Materia, 
Unit\`a di Como, Via Valleggio 11, 22100 Como, Italy}   
\address{$^{(2)}$Istituto Nazionale di Fisica Nucleare, 
Sezione di Milano, Via Celoria 16, 20133 Milano, Italy}   
\address{$^{(3)}$Center for Applied Mathematics and Theoretical 
Physics, University of Maribor, Krekova ul. 2, SI-2000 Maribor,
Slovenia}
\date{April 16, 2003}
\maketitle

\begin{abstract} 
We study both analytically and numerically the decay of fidelity 
of classical motion for integrable systems. We find that the
decay can exhibit two qualitatively different behaviors, namely an 
{\it algebraic decay}, that is due to the perturbation of the shape 
of the tori, or a {\it ballistic decay}, 
that is associated with perturbing the frequencies of the tori. The 
type of decay depends on initial conditions and on the shape 
of the perturbation but, for small enough perturbations, not on 
its size. We demonstrate numerically this 
general behavior for the cases of 
the twist map, the rectangular billiard, and the kicked 
rotor in the almost integrable regime.
\end{abstract} 

\begin{multicols}{2}
\narrowtext

\section{Introduction}

The study of the relation between classical and quantum dynamical 
chaos has greatly improved our understanding of the behavior of 
quantum systems. An issue which may be of interest in several 
situations is the stability of motion. In this respect,  
even though Liouville equation, which governs the evolution of 
classical distribution functions, is linear, the exponential sensitivity 
of classical trajectories with respect to perturbing the initial 
conditions leads to a strong dynamical instability,  
and characterizes classical chaos. 
Like the classical Liouville equation, the 
Schr\"odinger equation is also linear. However, the quantum evolution 
of states is stable and this qualitative difference is clearly 
apparent in the Loschmidt echo numerical experiments of Ref. \cite{arrow}. 
The problem of the stability of quantum motion under perturbations in 
the Hamiltonian has recently gained a renewed interest 
\cite{peres,jalabert,pastawski,jacquod,tomsovic,felix,zurek,PRE,cohen,prosen,lloyd,baowenli,mirlin,saraceno,heller}, also 
in connection with quantum computation \cite{prosenqc,simone,lloyd2}. 
The quantity of central interest in these investigations is the 
fidelity $f_q(t)$ (also called Loschmidt echo), which measures the 
accuracy to which a quantum state 
can be recovered by inverting, at time $t$, the dynamics with a 
perturbed Hamiltonian.
The main interest has been focused on classically chaotic systems
for which, besides numerical experiments, some theoretical tools are 
available, like random matrix theory and semiclassical methods. 
However, in the general case, the phase space structure is mixed 
with chaotic components and islands of stability.
If the motion starts inside an integrable island, then it very much 
resembles the motion in integrable systems.
Contrary to chaotic systems, which are dynamically unstable but 
structurally stable, integrable systems are 
dynamically stable but very sensitive to external perturbations. 
Therefore the analysis of the fidelity requires particular care and 
one may expect it to be dependent on initial conditions and on
the type of perturbation. 
Indeed, the decay of fidelity in integrable systems has been discussed 
in recent papers \cite{prosen,eckhardt,jacquod2}, and very different 
behaviors have been found. Jacquod {\it et al.} have shown the 
existence of a regime in which the quantum fidelity for 
classically integrable systems decays as a power law, with an 
anomalous exponent of purely quantum origin 
\cite{jacquod2}.  
Prosen and \v Znidari\v c have instead discussed a regime in which 
quantum fidelity exhibits a much faster Gaussian decay 
\cite{prosen}. Both regimes have 
been also discussed by Eckhardt in his analysis of the decay of 
classical fidelity \cite{eckhardt}, in which the problem 
of the evolution of classical phase space densities has been 
addressed for linearized flows. 

In the present paper, we discuss the behavior of fidelity for 
integrable classical systems. 
Besides being of interest on its own, our classical study will 
allow us to understand the main mechanisms for the fidelity decay 
and therefore will constitute a valuable 
reference point for the quantum analysis.
Here we show the existence of a critical border depending  
on the {\it shape} of the perturbation, which 
separates two different types of fidelity decay: a power law 
decay $\propto 1/t^n$, where $n$ is the dimension of the system, 
and a much faster decay of ballistic type.
We stress that the type of decay depends on initial 
conditions and on the shape of the perturbation but, for small 
enough perturbations, not on its strength.
We derive an analytical expression for the critical border and 
our theoretical results are confirmed by a numerical analysis on 
three different models: the twist map, the rectangular billiard and 
the kicked rotor. 

The outline of the paper is as follows: In Section II we develop a 
general theory for the decay of classical fidelity in integrable
systems. Section III demonstrates numerically the validity of our theory 
in two different typical examples of integrable systems, the 
twist map and the rectangular billiard, and in an almost integrable 
system, the kicked rotor.
Our conclusions are drawn in Section IV.
Finally, in the Appendix A, we discuss the long-time 
relaxation to equilibrium.  

\section{Theory}

The quantum fidelity is defined as the overlap at time $t$ of the
states $|\psi(t)\rangle$ and $|\psi_\epsilon(t)\rangle$,
obtained by the evolution of the same initial state 
$\left|\psi(0)\right>$
with the unperturbed Hamiltonian $H_0$ 
and the perturbed Hamiltonian $H_0+\epsilon V$, respectively. 
The fidelity is then given by
\begin{equation}
f_q(t)=\left|\left<\psi(t)|\psi_\epsilon(t)\right>\right|^2.
\end{equation}
This expression can be equivalently rewritten in terms of the Wigner 
functions as
\begin{equation}
f_q(t)=(2 \pi \hbar)^n
\int d^n {\bf q}~d^n{\bf p}~W_\epsilon({\bf q},{\bf p};t)~
W({\bf q},{\bf p};t),
\label{fidq}
\end{equation}
where $n$ is the number of degrees of freedom. Since the Wigner functions 
can be considered as the quantum analogues of the classical 
phase space densities, we define
the classical fidelity as
\begin{equation}
f(t)=\int d^n {\bf q}~d^n{\bf p}~\rho_\epsilon({\bf q},{\bf p};t)~
\rho({\bf q},{\bf p};t),
\label{fidc}
\end{equation}
where $\rho, \rho_\epsilon$ are the square normalized classical phase space 
densities 
($\int d^n {\bf q}~d^n{\bf p}~\rho^2=
\int d^n {\bf q}~d^n{\bf p}~\rho_\epsilon^2=1$).
We note that $f(t)$ is the classical limit of $f_q(t)$. 
As the density evolution is unitary in both classical and quantum mechanics, 
instead of evolving two densities forward in time and calculating their 
overlap, we may first evolve the initial density 
$\rho_0$ forward in time with the 
unperturbed Hamiltonian $H_0$, and then evolve this density backward
in time with the perturbed Hamiltonian $H_0+\epsilon V$. We denote 
the density obtained in such a way as $\rho_{2t}$. The fidelity is then 
given by the overlap of the density $\rho_{2t}$ 
with the initial density $\rho_0$:
\begin{equation}
f(t)=\int d^n {\bf q}~d^n{\bf p}~\rho_{2 t}({\bf q},{\bf p})~
\rho_0({\bf q},{\bf p}).
\label{fidc2}
\end{equation}
Such an approach is  more convenient for our discussion of the classical 
fidelity of integrable systems.

For the following discussion we assume that the perturbation of the 
integrable system is of the KAM type, namely that, for small 
enough perturbations $\epsilon V$, most of the tori of 
the system are only slightly deformed but not destroyed. 
Therefore for most of the tori the transformation 
from old action-angle variables ${\bf I},{\bf \Theta}$ to new ones 
${\bf I}^\prime,{\bf \Theta}^\prime$ is possible, in such a way that 
the new actions are constants of motion of the perturbed system. 
To the first order in the perturbation
strength $\epsilon$ the transformation can be written as
\begin{eqnarray}
{\bf \Theta}^\prime  &=& {\bf \Theta} + 
\epsilon{\bf f}({\bf I}, {\bf \Theta}),
\label{eq:transtheta}\\
{\bf I}^\prime &=& {\bf I} + 
\epsilon{\bf g}({\bf I}, {\bf \Theta}).
\label{eq:transi}
\end{eqnarray}

After the forward unperturbed evolution up to time 
$t$ we have 
\begin{equation}
{\bf \Theta}_{t}={\bf \Theta}_0+{\bf \Omega}({\bf I}) t.
\end{equation}
Then we perform a backward evolution of the perturbed system from 
time $t$ to time $2t$, getting
\begin{equation}
{\bf \Theta}_{2t}^\prime = {\bf \Theta}_t^\prime - 
{\bf \Omega}^\prime({\bf I}^\prime) t.
\end{equation}
The frequency vectors ${\bf \Omega}$ and ${\bf \Omega}^\prime$ 
characterize the linear (in time) evolution of the angle variables 
in integrable systems. 
Expressed in the original action-angle variables, the overall evolution 
can be written as
\begin{equation}
{\bf \Theta}_{2t}={\bf \Theta}_0+\left({\bf \Omega}({\bf I})
-{\bf \Omega}^\prime({\bf I}^\prime)\right)t + {\cal O} (\epsilon)
\label{eq:theta2t},
\end{equation}
where the error term ${\cal O} (\epsilon)$ is due to the change from one 
set of variables to the other at time $t$ and the reverse change at time $2t$. 
Since the perturbation is small, we may write the change in frequency as
\begin{equation}
{\bf \Omega}^\prime({\bf I}^\prime)={\bf \Omega}({\bf I})+
\Delta {\bf \Omega}({\bf I})
+\frac{\partial {\bf \Omega}}{\partial {\bf I}} ({\bf I}^\prime-{\bf I})
+{\cal O}(\epsilon^2) \label{eq:omegaexpand},
\end{equation}
where $\Delta {\bf \Omega}\equiv {\bf \Omega}^\prime({\bf I})-
{\bf \Omega}({\bf I})$ denotes the change of frequencies on the 
unperturbed torus ${\bf I}$ and 
${\bf I}^\prime-{\bf I}$ gives the change of the action 
variables caused by the perturbation (written in Eq.~(\ref{eq:transi})
to the first order in $\epsilon$). 

If we consider the angle variables to be uniformly distributed at the time 
$t$ at which the motion is inverted, we may introduce the distribution
\begin{equation}
W_{\bf I}\left(\frac{\Delta {\bf I}}{\epsilon}\right)=
\frac{1}{(2 \pi)^n}\int d^n {\bf \Theta}~
\delta\left[
\frac{\Delta {\bf I}-({\bf I}^\prime({\bf I},{\bf \Theta})
-{\bf I}))}{\epsilon}
\right], 
\label{eq:probaction}
\end{equation}
which gives the probability density for the transition from the torus 
characterized by the action variables ${\bf I}$ in the unperturbed
coordinates to the torus with action variables ${\bf I}^\prime$ in the 
perturbed coordinates. 
The $\epsilon$ scaling has been chosen in order for the function 
$W_{\bf I}$ itself not to depend on $\epsilon$ 
in the linear approximation.

In the generic case, the motion on a torus 
is ergodic. It is, however, not random and therefore, in order for the 
equation (\ref{eq:probaction}) to well describe transitions between the tori, 
one needs to consider an ensemble of tori in the vicinity of the chosen 
actions ${\bf I}$, as only for an ensemble of tori with different 
frequencies we can expect the angle variables to be uniformly distributed 
after a sufficiently long time $t$. 
Indeed, the spread of frequencies given by 
$\delta {\bf \Omega}=\frac{\partial {\bf \Omega}}{\partial 
{\bf I}} \delta {\bf I}$ translates into the spread of angle 
variables $\delta {\bf \Theta}=\delta {\bf \Omega}~t$. 
The time for this spread in the angle variable to become comparable to 
$2 \pi$ is 
\begin{equation}
t_\theta\approx \frac{2 \pi}{\left(
\frac{\partial \Omega}{\partial I}
\right)\nu_{I}},
\label{ttheta}
\end{equation}
where $\nu_I$ is the characteristic width of the initial phase space 
density distribution $\rho_0$ along the action direction
(to simplify writing, we have given Eq.~(\ref{ttheta}) for the 
one-dimensional case).
Thus, our theory based on Eq.~(\ref{eq:probaction}) is valid for times 
$t>t_\theta$. 

Since the final angle variables, after the forward and backward evolutions, 
depend on the actions of the perturbed system, we compute the distribution 
of the angle variables from the distribution of the perturbed actions as
\begin{equation}
P_{{\bf I}}({\bf \Theta}_{2t}-{\bf \Theta}_{0};t)=
W_{{\bf I}}\left(\frac{{\bf I}^\prime-
{\bf I}}{\epsilon}\right)
\left|\frac{\partial (({\bf I}^\prime-
{\bf I})/\epsilon)}{\partial ({\bf \Theta}_{2t}-
{\bf \Theta}_{0})}\right|.
\end{equation}
Using the expressions (\ref{eq:theta2t}) and (\ref{eq:omegaexpand}), 
we obtain  
$$
P_{{\bf I}}({\bf \Theta}_{2t}-{\bf \Theta}_{0};t)=
\frac{1}{(\epsilon t)^n}
\left|\frac{\partial {\bf \Omega}}{\partial {\bf I}}\right|^{-1}\times
$$
\begin{equation}
W_{{\bf I}}\left(
\left[\frac{\partial {\bf \Omega}}{\partial {\bf I}}\right]^{-1}
\left(\frac{{\bf \Theta}_{0}- {\bf \Theta}_{2t}+{\cal O} (\epsilon)}
{\epsilon t}-\frac{\Delta {\bf \Omega}}{\epsilon}\right)\right).
\label{eq:main}
\end{equation}
This expression is the kernel 
for the combined forward and backward evolution 
of the phase space densities. 

We assume that the width $\nu_I$ along the action direction
of the initial density $\rho_0$ is 
much larger than the change of the action variable 
induced by the perturbation, that is 
\begin{equation}
\nu_{I}\gg\epsilon\max_{I,\Theta}
|g(I,\Theta)|
\label{nuI}
\end{equation}
(to simplify writing we have given the condition  
(\ref{nuI}) for the one-dimensional case).
Therefore the effects of the forward and backward evolutions are
felt mainly in the change of the angles variable. This means that 
the evolution from the initial phase space density $\rho_0$ to 
$\rho_{2t}$ is given, up to corrections of order $\epsilon$, by
\begin{equation}
\rho_{2t}({\bf I},{\bf \Theta})=\int d^n{\bf \Theta}^\prime~ 
P_{{\bf I}}({\bf \Theta}^\prime-{\bf \Theta};t)~
\rho_0({\bf I},{\bf \Theta}^\prime).
\label{rho2t}
\end{equation}
The fidelity $f(t)$ can then be computed by inserting 
$\rho_{2t}$ into Eq.(\ref{fidc2}). 

The kernel $P_{{\bf I}}({\bf \Theta}^\prime-{\bf \Theta})$ is  
stretched linearly in time, while at the same time it  
moves ballistically (linearly with time) with 
velocity $\Delta {\bf \Omega}$. 
Under the assumption that the perturbation of the 
shape of the tori is not divergent
(as it is the case for most of the tori in a KAM regime), the distribution
$W_{{\bf I}}({\bf I}/\epsilon)$ has a bounded support 
which is determined by the 
change of the shape of the tori due to the perturbation. 
We can see from Eq.~(\ref{eq:main}) that at long times the 
argument of the function $W_{\bf I}$ is given by 
$-[\partial {\bf \Omega}/\partial {\bf I}]^{-1}
(\Delta {\bf \Omega}/\epsilon)$. 
Therefore the long time behavior of $P_{\bf I}$ depends on whether 
the value of
$-[\partial {\bf \Omega}/\partial {\bf I}]^{-1}
(\Delta {\bf \Omega}/\epsilon)$ falls within the support of
$W_{\bf I}$ or not. In the first case,   
$\tilde{W}_{\bf I}\equiv 
W_{\bf I}(-[\partial {\bf \Omega}/\partial {\bf I}]^{-1}
(\Delta {\bf \Omega}/\epsilon))$ is different from zero
and therefore the kernel $P_{\bf I}$ drops $\propto 1/t^n$.
In the latter case, $\tilde{W}_{\bf I}=0$, and therefore $P_{\bf I}$ 
drops ballistically. The transition between these two 
regimes is determined by the equality  
\begin{equation}
\frac{\Delta {\bf I}_s}{\epsilon}=
-\left[\frac{\partial {\bf \Omega}}{\partial {\bf I}}\right]^{-1}
\frac{\Delta {\bf \Omega}}{\epsilon},
\label{DeltaIs}
\end{equation}
where $\Delta {\bf I}_s/\epsilon$ are the coordinates of the border 
of the support of $W_{\bf I}$.

We can therefore draw the following conclusions: If the 
perturbation of a classical integrable system is such that the primary 
effect is the change of the shape of the tori, then the expected decay 
of fidelity is $\propto 1/t^n$. On the contrary,
if the change of the frequencies of the tori is the dominant 
effect, then we expect a ballistic decay of fidelity, 
that is the center of mass motion of the phase space densities
after the forward and backward evolutions is responsible for a  
drastic drop of fidelity. Such a decay takes place 
as soon as the centers of mass of 
the densities $\rho_{2t}$ and $\rho_0$ are separated in the 
angle variables ${\bf \Theta}$ by more than their characteristic 
width ${\bf \nu}_{\bf \Theta}$. 
As it can be seen from Eq.~(\ref{fidc2}), the exact form of the 
fidelity drop in the ballistic regime depends on the 
tails of the initial distribution $\rho_0$: for instance, a 
Gaussian tail gives a Gaussian decay of fidelity, whereas a 
sharp border induces a sharp drop to zero of fidelity.   
Finally, it is important to stress that the type of 
decay, power law or ballistic, depends on initial conditions and 
on the shape of the perturbation. However, it does not depend on 
the strength of the perturbation, provided that it is sufficiently 
small. 
Indeed, Eq.~(\ref{DeltaIs}) shows 
that $\Delta {\bf I}_s/\epsilon$ is $\epsilon$-independent
(to the first order in $\epsilon$), 
since $\Delta {\bf \Omega}\propto \epsilon$. 

\section{Numerical demonstration}
As a first example we consider the perturbed twist map, defined by 
\begin{eqnarray}
I_{t+1}&=&I_{t}+\epsilon \cos(\alpha) 
\sin(\Theta_{t})\label{eq:skew1}, \nonumber \\
\Theta_{t+1}&=&\Theta_{t}+I_{t+1}+
\epsilon \sin(\alpha)\sin(I_{t+1}),
\label{eq:skew2}
\end{eqnarray}
where the angle $\alpha$ determines the mixture between purely 
perturbing the shape of the tori ($\alpha=0$) or purely changing 
their frequencies ($\alpha=\pi/2$).
This parametrization allows us to change the type of the perturbation 
without changing its overall magnitude.
The change of frequency associated with the perturbation is given by 
\begin{equation}
\Delta \Omega=\epsilon \sin(\alpha) \sin(I).
\end{equation}
The conserved action variable of the $\epsilon$-perturbed system 
is, to the first order (in $\epsilon$) approximation, given by
\begin{equation}
I^\prime=I+\epsilon \cos(\alpha) 
\frac{1}{2 \sin(I/2)} \cos\left(\Theta-\frac{I}{2}\right).
\end{equation}
Indeed, inserting this expression into the mapping 
(\ref{eq:skew2}), one can easily verify that $I_{t+1}=I_t$. 
The transition probability function 
$W_I$ for this system can thus be obtained by means of 
Eq.~(\ref{eq:probaction}):
$$
W_{I}
\left(\Delta I/\epsilon\right)=
$$
\begin{equation}
\frac{1}{2 \pi} \int d \Theta\ \delta
\left[\frac{\Delta I}{\epsilon} 
-\frac{\cos(\alpha)}{2 \sin(I/2)}
\cos\left(\Theta-\frac{I}{2}\right)\right],
\end{equation}
which gives
\begin{equation}
W_{I}
\left(\frac{\Delta I}{\epsilon}\right)=
\frac{1}{\pi \sqrt{\left(\frac{\cos(\alpha)}{2 \sin(I/2)}\right)^2-
\left( \Delta I/\epsilon\right)^2}}.
\label{eq:distskew}
\end{equation}
The range of the support of the distribution $W_I$ is between 
$\pm {\cos(\alpha)}/{(2 \sin(I/2))}$.
The critical value $\alpha_c$ for which 
$\Delta I_s/\epsilon=
-(\partial \Omega/\partial I)^{-1} \Delta \Omega/\epsilon$ 
is therefore determined by
\begin{equation}
\tan(\alpha_c)=\frac{1}{2 \sin(I) \sin(I/2)}. 
\label{eq:critskew}
\end{equation}

\begin{figure}
\includegraphics[width=8cm]{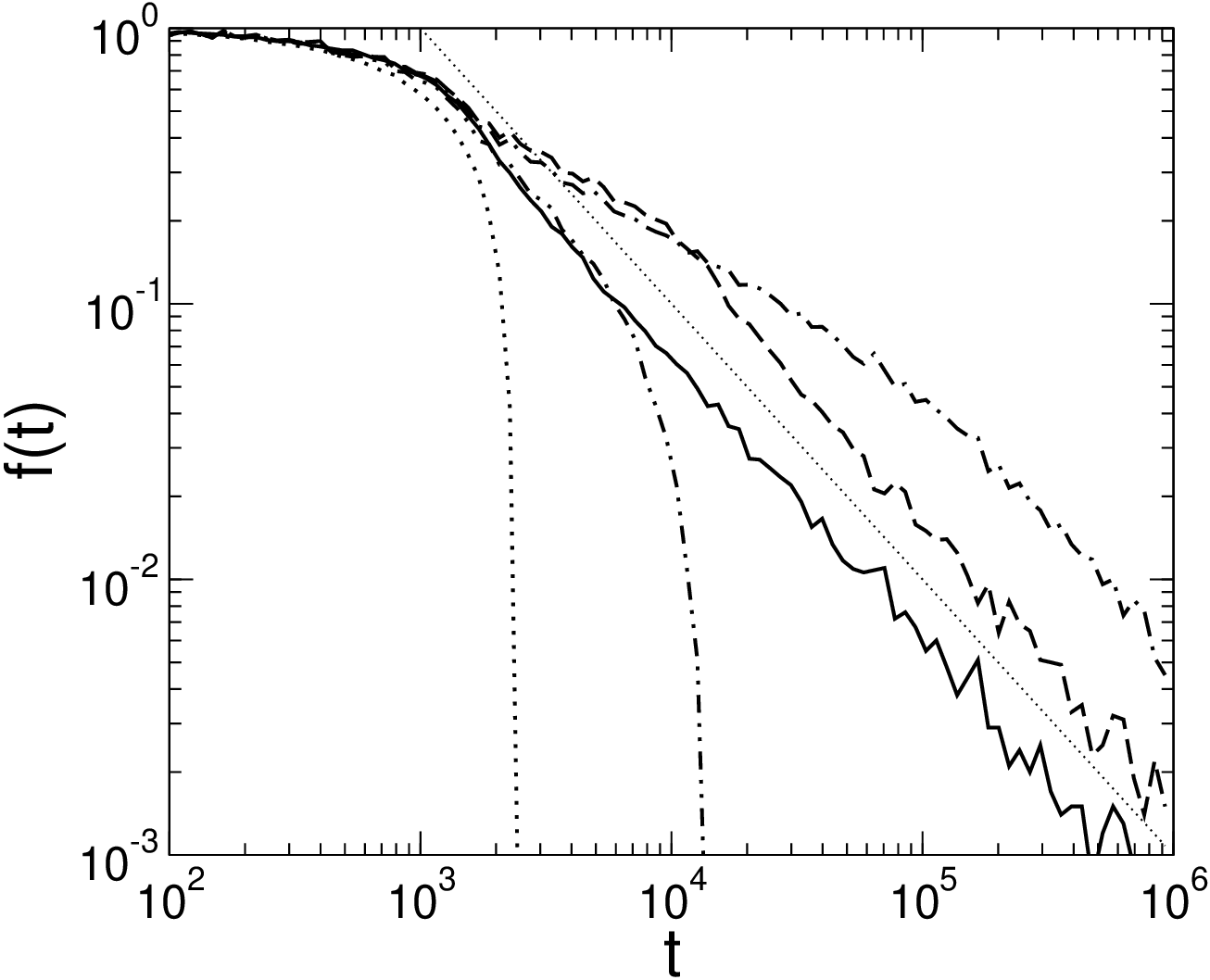}
\caption{Fidelity decay for the twist map 
at various values of the parameter $\alpha=$ 0 (full line), 0.8 (dashed), 
0.892 (dot-dashed), 1.0 (dot-dot-dashed) and $\pi/2$ (dotted). 
The $\propto 1/t$ decay is shown as a thin dotted line }
\label{fig:skew}
\end{figure}

In Fig.~\ref{fig:skew} we show the numerically computed
behavior of fidelity for this system
as a function of time for various values of the parameter $\alpha$. 
We take as initial phase space density a rectangle centered around the 
point $(\Theta=\pi,I=1)$ with sides of length $\nu_\Theta= 2
\times 10^{-3},\ \nu_I =2\times 10^{-2}$. The perturbation
strength is $\epsilon=10^{-6}$. 
To compute fidelity, we follow the evolution
of $N=10^4$ trajectories, which at $t=0$ are uniformly distributed
inside the above rectangle. The fidelity
is then given by the percentage of trajectories that
return back to this region after the forward and backward 
evolutions. 
In all cases we 
observe an initial plateau during which the fidelity does not decay
appreciably. This plateau persists until the time $t_p$ at which 
the width of the kernel (\ref{eq:main}) or the shift of its 
center become comparable to the 
width $\nu_\Theta$ of the phase space density along the 
angle variable. In either case this time is 
\begin{equation}
t_p\propto \frac{\nu_\Theta}{\epsilon}.
\label{tplateau}
\end{equation}
According to Eq.~(\ref{DeltaIs}), we expect the behavior 
to change from algebraic decay to ballistic one
at the value of the parameter $\alpha=\alpha_c$.
For the chosen initial conditions, Eq.~(\ref{eq:critskew})
gives $\alpha_c\approx 0.892$. 
Indeed, the change from an algebraic fidelity decay $f(t)\propto 1/t$ 
when $\alpha<\alpha_c$ to a sharp drop of fidelity
when $\alpha>\alpha_c$ is clearly seen in Fig.~\ref{fig:skew}. 
 
An interesting feature is that, approaching the 
critical value $\alpha_c$, we observe 
that the fidelity decay, power law or ballistic, sets in after 
longer and longer times. 
This fact has a clear explanation: The value  
$\tilde{W}_{\bf I}=W_I(-[\partial {\bf \Omega}/\partial  {\bf I}]^{-1} 
(\Delta {\bf \Omega}/\epsilon))$,
which determines the long time behavior of the evolution 
kernel (\ref{eq:main}), diverges close to the critical value 
$\alpha=\alpha_c$ (see Eq.~(\ref{eq:distskew})). 
When the fidelity decay is power law, we have
$P_{\bf I}=c/t^n$, with the constant $c\propto W_{\bf I}$.
Since $c$ becomes larger and larger close to the critical point, 
the fidelity decay must be postponed to longer and longer times.   
On the other hand, when the long time fidelity decay 
is ballistic, we can see 
from Eq.~(\ref{eq:main}) that the argument of the function 
$W_{\bf I}$ goes outside the support of $W_{\bf I}$ after 
a time that becomes longer close to the critical point. 
Only after this time the fidelity drops off.
Of course the decay cannot be postponed indefinitely
since the exact condition (\ref{eq:critskew}) can be satisfied 
only for a single torus, while we always deal with a family of tori 
upon which the initial phase space density $\rho_0$ rests. 

We also checked numerically that, provided that the perturbation 
is much smaller than the characteristic widths $\nu_I$ of the 
initial density (that is, the requirement (\ref{nuI}) is fulfilled),
the type of behavior does not alter 
with changing the actual size of the perturbation $\epsilon$, as 
it is expected from our theory. We simply rescale the time 
$t_p\propto 1/\epsilon$ after which the fidelity decay starts,
in agreement with Eq.~(\ref{tplateau}). 

To illustrate the fidelity decay in integrable systems with 
more than one degree of freedom, we consider the following 
system: 
\begin{equation}
H(I_1,I_2,\Theta_1,\Theta_2)= H_0(I_1,I_2)+ 
\epsilon V(I_1,I_2,\Theta_1,\Theta_2),
\label{Hrectangle}
\end{equation}
where the unperturbed Hamiltonian 
\begin{equation}
H_0=\frac{\alpha_1}{2}I_1^2+\frac{\alpha_2}{2}I_2^2
\end{equation}
describes the motion of a particle bouncing elastically 
inside a rectangular billiard and the perturbation is 
given by  
\begin{equation}
V=\cos(\beta) \cos(\Theta_1) \cos(\Theta_2)
+\sin(\beta) I_1 I_2\nonumber.
\end{equation}
Again, depending on the value of the parameter $\beta$, the 
perturbation mainly affects either the shape of the tori or 
their frequencies.
We use the first order perturbation theory of Hamiltonian systems 
(see, e.g., Ref.~\cite{ref:pert}) to determine
the effects of the perturbation. 
What we need to find is a set of action-angle
coordinates such that, to the first order in the perturbation 
strength $\epsilon$, the Hamiltonian (\ref{Hrectangle})
in these new coordinates can be written as a function 
of only the new actions, namely 
\begin{equation}
H(I_1,I_2,\Theta_1,\Theta_2)=
H^\prime(I_1^\prime,I_2^\prime)+{\cal O}(\epsilon^2).
\label{eq:hamilaction}
\end{equation}
Introducing the generating function 
$$
G(I_1^\prime,I_2^\prime,\Theta_1,\Theta_2)=
I_1^\prime\Theta_1+I_2^\prime\Theta_2
$$
\begin{equation}
-\frac{\epsilon\cos\beta}{2}\left[
\frac{\sin(\Theta_1+\Theta_2)}{\alpha_1 I_1^\prime +
\alpha_2 I_2^\prime}+
\frac{\sin(\Theta_1-\Theta_2)}{\alpha_1 I_1^\prime -
\alpha_2 I_2^\prime}\right],
\end{equation}
we get 
$$
I_1=\frac{\partial G}{\partial \theta_1}=
I_1^{\prime}-\frac{\epsilon \cos(\beta)}{2}
\left[\frac{1}{\alpha_1 I_1+\alpha_2 I_2}
\cos(\Theta_1+\Theta_2)\right.
$$
\begin{equation}
\left.+\frac{1}{\alpha_1 I_1-\alpha_2 I_2}
\cos(\Theta_1-\Theta_2)
\right],
\end{equation}
$$
I_2=\frac{\partial G}{\partial \theta_2}=
I_2^{\prime}-\frac{\epsilon \cos(\beta)}{2}
\left[\frac{1}{\alpha_1 I_1+\alpha_2 I_2}
\cos(\Theta_1+\Theta_2) \nonumber \right. 
$$
\begin{equation}
\left. -\frac{1}{\alpha_1 I_1-\alpha_2 I_2}
\cos(\Theta_1-\Theta_2) \right].
\end{equation}
Substituting the above expressions into 
the Hamiltonian (\ref{Hrectangle}), we get
\begin{equation}
H^\prime(I_1^\prime,I_2^\prime)=
\frac{\alpha_1^2}{2}{I_1^\prime}^2+
\frac{\alpha_2^2}{2}{I_2^\prime}^2+
\epsilon\sin(\beta) I_1^\prime I_2^\prime.
\end{equation}
The new frequencies are then given by  
\begin{equation}
\Omega_1^\prime=
\frac{\partial H^\prime}{\partial I_1^\prime}=
\alpha_1 I_1^\prime+ \epsilon 
\sin(\beta) I_2^\prime,
\end{equation}
\begin{equation}
\Omega_2^\prime=
\frac{\partial H^\prime}{\partial I_2^\prime}=
\alpha_2 I_2^\prime+ \epsilon 
\sin(\beta) I_1^\prime.
\end{equation}
Thus the frequencies changes read as follows: 
\begin{equation}
\Delta \Omega_1=\epsilon\sin(\beta) I_2,
\end{equation}
\begin{equation}
\Delta \Omega_2=\epsilon\sin(\beta) I_1.
\end{equation} 
As in the previous example, 
the above expressions allow us to find the transition probability 
function 
\begin{eqnarray}
W_{{\bf I}}(\Delta I_1/\epsilon,\Delta I_2/\epsilon)\nonumber
=\quad\\
\nonumber\\
\frac{2}{\pi^2}
\frac{1}{\sqrt{\left(\frac{\cos(\beta)}{\alpha_1 I_1+\alpha_2 I_2}\right)^2
-\left(\frac{\Delta I_1+\Delta I_2}{\epsilon}\right)^2}}\times\nonumber\\
\nonumber \\
\frac{1}{\sqrt{\left(\frac{\cos(\beta)}{\alpha_1 I_1-\alpha_2 I_2}\right)^2
-\left(\frac{\Delta I_1-\Delta I_2}{\epsilon}\right)^2}}.
\end{eqnarray}
It can be seen that the support for the distribution
$W_{{\bf I}}$ is the rectangle 
\begin{equation}
|U|<\frac{\cos(\beta)}{\alpha_1 I_1+\alpha_2 I_2},
\quad
|V|<\frac{\cos(\beta)}{\alpha_1 I_1-\alpha_2 I_2},
\end{equation} 
where 
$U=(\Delta I_1+\Delta I_2)/\epsilon$ and 
$V=(\Delta I_1-\Delta I_2)/\epsilon$.

\begin{figure}
\includegraphics[width=8cm]{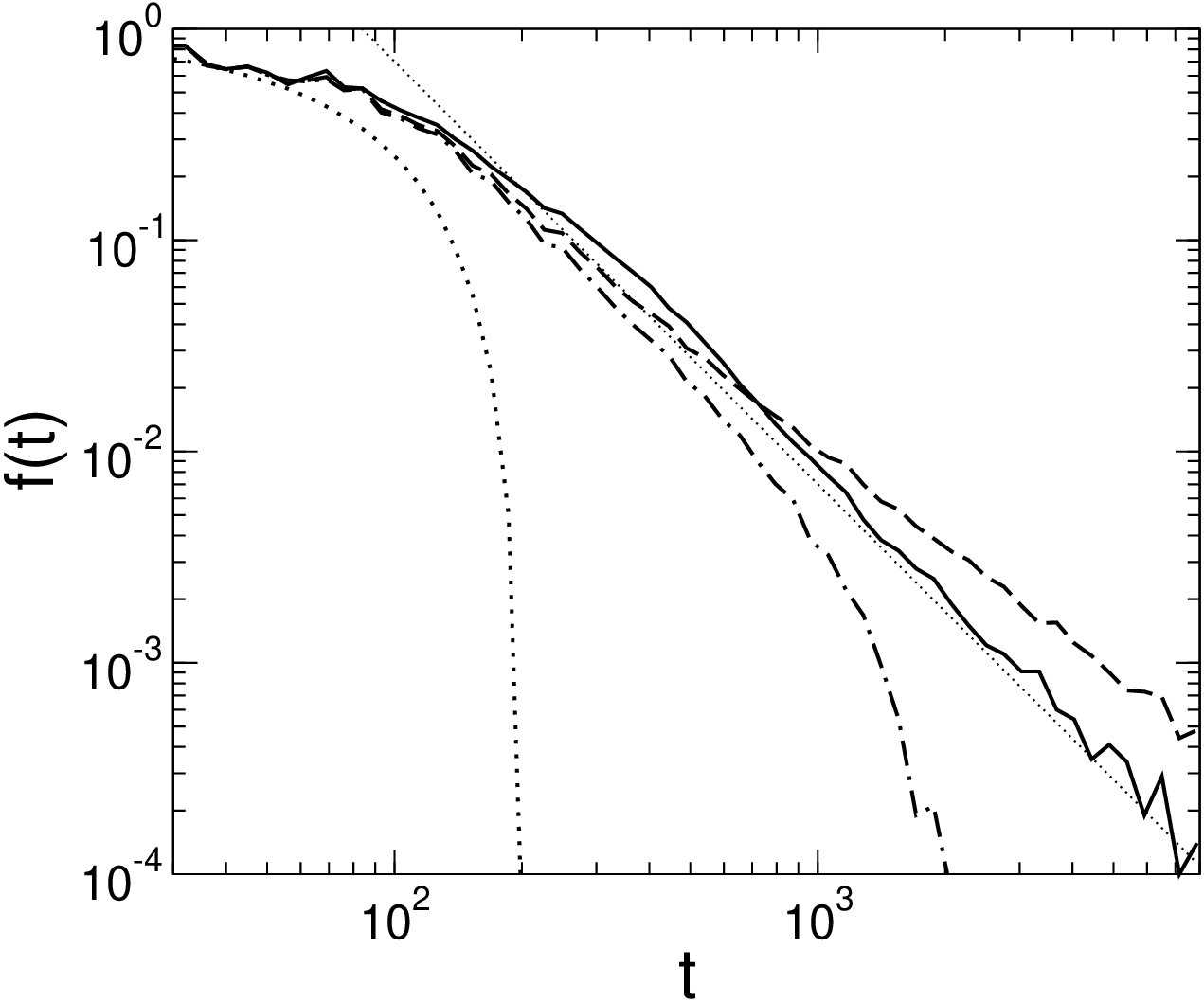}
\caption{Fidelity decay for the rectangular billiard 
for various values of the parameter $\beta=$ 0 (full line),  
0.232 (dashed), 0.3 (dot-dashed)
and $\pi/2$ (dotted). The $\propto 1/t^2$ decay is 
shown as a thin dotted line.}
\label{fig:rect}
\end{figure}

In Fig.~\ref{fig:rect}
we show the decay of fidelity for this system for various values
of $\beta$. 
The parameters
of the system have been chosen as follows: $\alpha_1=(\sqrt{5}+1)/2$, 
$\alpha_2=1$. In all cases
the initial phase space density is a hyper-rectangle centered around
$I_1=1$, $I_2=1$, $\Theta_1=1$ and $\Theta_2=1$ with all sides
of length $\nu_{I_1}=\nu_{I_1}=
\nu_{\Theta_1}=\nu_{\Theta_2}=0.02$. 
The perturbation parameter is $\epsilon=3\times 10^{-4}$ and 
the number of trajectories $N=10^5$. 
For the above initial conditions and parameters $\alpha_1$,
$\alpha_2$, the critical value $\beta_c$ which separates the 
power law and the ballistic fidelity decay is determined by the 
equality (\ref{DeltaIs}) in the direction of the $U$ variable. 
Indeed, when $\beta$ increases, $-[\partial {\bf \Omega}/
\partial {\bf I}]^{-1}(\Delta {\bf \Omega}/\epsilon)$ 
goes outside the support of $W_{\bf I}$ at first along 
this direction. This gives  
\begin{equation} 
-\left(\frac{\Delta {\bf I}_s}{\epsilon}\right)_U=
\frac{\cos(\beta_c)}{\alpha_1 I_1 + \alpha_2 I_2}=
\left[
\left(\frac{\partial {\bf \Omega}}{\partial {\bf I}}\right)^{-1} 
\Delta {\bf \Omega}
\right]_U,
\label{betaU} 
\end{equation}
where the right-hand side is the $U$-component of the vector  
\begin{eqnarray}
\left(
\frac{\partial {\bf \Omega}}{\partial {\bf I}}
\right)^{-1}
\Delta {\bf \Omega}
=\left(
\begin{array}{cc} 
\alpha_2^{-1} & 0 \\
0 & \alpha_1^{-1}
\end{array}
\right) 
\left(
\begin{array}{c} 
\sin (\beta_c)I_2\\ 
\sin (\beta_c)I_1\\ 
\end{array}
\right).
\end{eqnarray}
Therefore we get
\begin{equation}
\tan(\beta_c)=\frac{\alpha_1\alpha_2}{(\alpha_1 I_1+\alpha_2 I_2)^2}.
\end{equation}
Substituting the chosen values of $I_1,I_2,\alpha_1,$ and $\alpha_2$,
we find that the critical value is equal to $\beta_c\approx0.232$. 
This theoretical expectation is confirmed 
by the numerical data of Fig.~\ref{fig:rect}, which show a 
crossover from a power law fidelity decay (for $\beta<\beta_c$) 
to a ballistic decay (for $\beta>\beta_c$).  
The results are very similar to the case of the twist map, 
including the fact that, close to the critical value $\beta=\beta_c$,
the decay is postponed to longer times. 
Is should be stressed that the algebraic fidelity decay, differently 
from the twist map case, is now inversely proportional to the 
square of the time, in agreement with our theoretical expectation
for a two-dimensional system (see Eq.~(\ref{eq:main})).

\begin{figure}
\includegraphics[width=8cm]{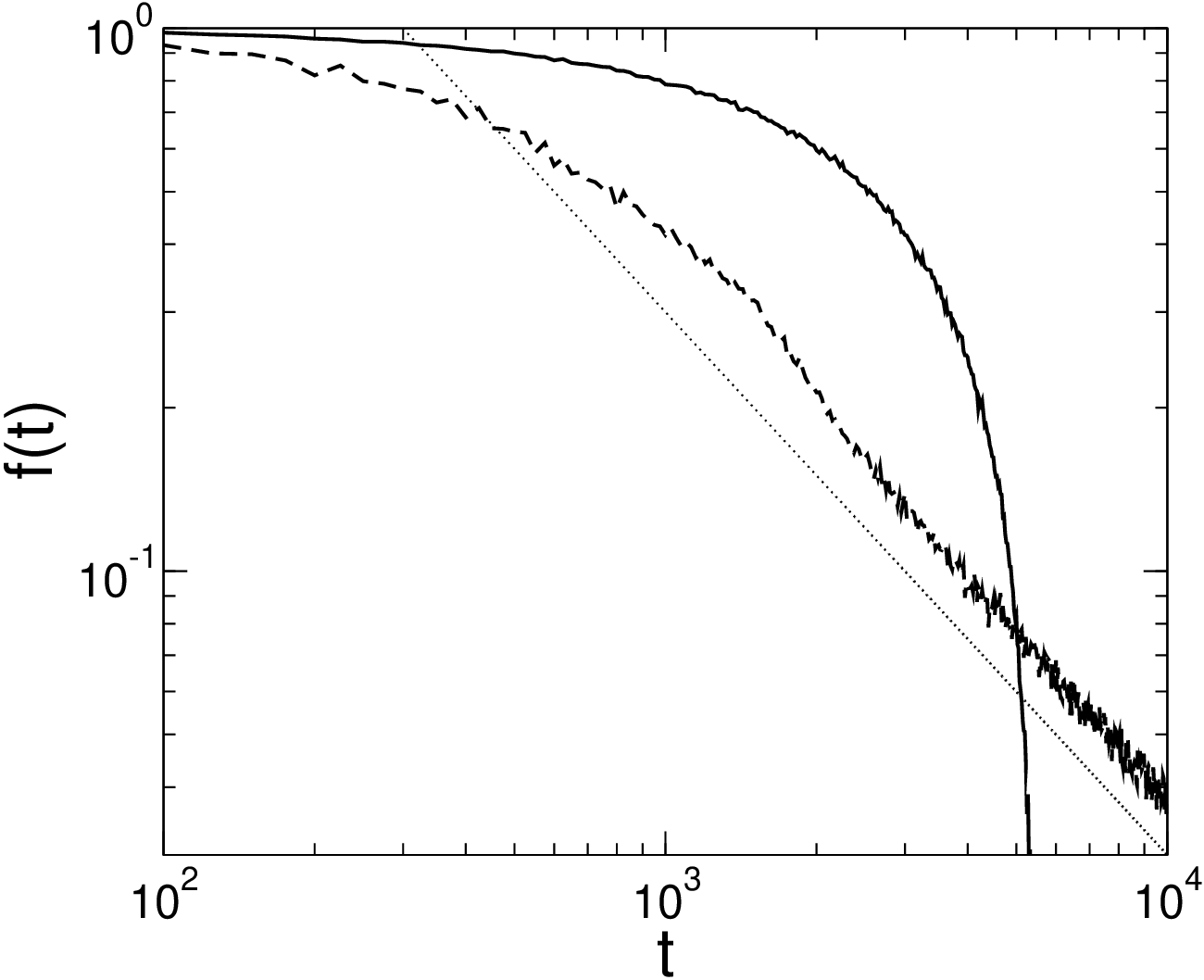}
\caption{Fidelity decay for the kicked rotor map with $K=0.3$,
$\epsilon=K^\prime-K=10^{-5}$, and $N=10^4$ trajectories,
using two different initial phase space 
densities, one centered at the point $(\theta=\pi,I=0.2)$
(full line) and the other centered at $(\theta=\pi,I=1.2)$ 
(dashed line). In the first case the fidelity decays ballistically, 
in the latter case inversely proportional to time. Both initial densities 
are $0.02$ wide in the $\Theta$ and $I$ directions. 
The $\propto 1/t$ curve is shown as a thin dotted line.}
\label{fig:rotor}
\end{figure}

As a last numerical example, we consider the kicked rotor
map that is given by
\begin{equation}
I_{t+1}=I_{t}+K \sin(\Theta_{t})\label{eq:rot1},
\end{equation}
\begin{equation}
\Theta_{t+1}=\Theta_{t}+I_{t+1}\label{eq:rot2}.
\end{equation}
As it is known, for $K\ll 1$ the system is almost integrable,
namely its phase space is dominated by invariant tori.
There is a stable fixed point at $(\Theta=\pi,I=0)$ and a 
separatrix which divides the phase space into two regions:
a section of librational motion around the stable fixed 
point inside the separatrix, and a section of 
rotational motion outside the separatrix.
We perturb the system by varying $K\to K^\prime =K+\epsilon$. 
The important point is that the type of the perturbation chosen strongly
affects the frequencies of the tori in the librational section, while
it mainly perturbs the shape of the rotational tori. Therefore 
the same system and perturbation should lead to two completely different 
types of fidelity decays, power law or ballistic, depending
on the choice of the initial conditions.
Fig.~\ref{fig:rotor} confirms this expectation:
if the initial density $\rho_0$ is inside the separatrix the 
fidelity decay is ballistic, otherwise it is power law.  

\section{Conclusions and outlook}

In this work we have studied the decay of the fidelity of classical
motion for integrable systems. 
Our main result is the following: for small enough perturbations,
the type of the decay of fidelity for integrable systems depends not 
on the strength of the perturbation but on its shape and on 
initial conditions. 
More precisely, the fidelity exhibits two 
completely different behaviors, namely an algebraic decay 
if the perturbation mainly affects the shape of the tori, and a faster, 
ballistic decay, if the main effect of the perturbation is to change 
the frequencies of the tori. We have also given clear numerical 
demonstrations of the transition between the two types of behaviors, 
induced by changing the shape of the perturbation or  
the initial conditions.

This result poses interesting questions with respect 
to the quantum mechanical picture.
Due to the correspondence principle, there should exist regimes 
where both types of decay may be observed. It is however 
expected that, for small perturbations, quantum mechanics
would favor the ballistic type decay, as demonstrated in 
\cite{prosen}. Indeed the algebraic
decay is due to the transitions between tori which,
for small perturbations, are suppressed in quantum mechanics, 
due to tori quantization and subsequent gaps between them. The
classical-quantum correspondence will be the topic of further studies. 

This work was supported in part by the EC RTN contract 
HPRN-CT-2000-0156, the NSA and ARDA under 
ARO contract No. DAAD19-02-1-0086, the
PA INFM ``Weak chaos: Theory and applications'', 
and the PRIN 2002 ``Fault tolerance, control and stability in 
quantum information precessing''.

\appendix
\section{Asymptotic behavior}

The results of the previous Sections do not tackle the asymptotic 
decay of fidelity for integrable systems \cite{gregor}. 
Indeed, we neglected the contributions to the 
evolution kernel (\ref{eq:main}) that stem from the fact that the 
angle variables are cyclic. This
means that, after a time which is $\propto{2\pi/\epsilon}$,
we need to take into account the contributions to (\ref{eq:main})
not only at ${\bf \Theta}_{2t} - {\bf \Theta}_0$ but also 
at all ${\bf \Theta}_{2t} - {\bf \Theta}_0+ 2\pi {\bf k}$, where 
${\bf k}$ is a vector of integer numbers.

We limit ourselves to the case of a single torus.
Of course the fidelity $f(t)$ is strictly zero for a single 
torus, and therefore it should be understood that we take the 
limits 
\begin{equation}
\epsilon\to 0,\;\;
{\bf \nu}_{\bf I}\to 0,\;\;
{\rm with}\;\;
\frac{\epsilon}{{\bf \nu}_{\bf I}}={\rm constant}\ll 1.
\end{equation}
Let us consider the initial density $\rho_0({\bf \Theta})$ to be 
defined on the whole ${\bf \Theta}$ space (without $2\pi$ periodicity), 
while the kernel $K_{\bf I}({\bf \Theta};t)$ is defined as the periodic
function obtained from the original kernel:
\begin{equation}
K_{\bf I}({\bf \Theta};t)=\sum_{\bf k} 
P_{\bf I}({\bf \Theta}-2 \pi {\bf k};t) 
\label{eq:perker}
\end{equation}
Assuming that the initial density is square normalized, the fidelity 
can be written as
\begin{equation}
f(t)=\int d^n {\bf \Theta} \rho_0^*({\bf \Theta}) \rho_{2t}({\bf \Theta})=
\int d^n {\bf \Phi} \tilde \rho_0^*({\bf \Phi}) \tilde \rho_{2t}({\bf \Phi}),
\label{eq:fidfour}
\end{equation}
where $\ \tilde{}\ $ denotes the Fourier transform:
\begin{equation}
\tilde \rho({\bf \Phi})=\frac{1}{\sqrt{2 \pi}} \int d^n {\bf \Theta}
\rho({\bf \Theta}) \exp(-i {\bf \Phi}{\bf \Theta}).
\end{equation}
Since
\begin{equation}
\rho_{2t}({\bf \Theta})=\int d^n {\bf \Theta^\prime } \rho_0({\bf \Theta^\prime})
K_{\bf I}( {\bf \Theta}-{\bf \Theta^\prime };t),
\end{equation}
in the Fourier picture this becomes
\begin{equation}
\tilde \rho_{2t}({\bf \Phi})=\tilde \rho_0({\bf \Phi})
\tilde K_{\bf I}({\bf \Phi};t).
\label{eq:evolfour}
\end{equation}
We may write the original kernel (\ref{eq:main}) in the simplified form
\begin{equation}
P_{\bf I}({\bf \Theta};t)=1/t^n 
p_{\bf I}({\bf \Theta}/t + {\bf \Gamma} ),
\end{equation}
where $p_{\bf I}({\bf y})=\epsilon^{-n} W_{\bf I}\left(
-\left[\frac{\partial {\bf \Omega}}{\partial {\bf I}}\right]^{-1} 
\frac{{\bf y}}{\epsilon}
\right)
\left|\frac{\partial {\bf \Omega}}{\partial {\bf I}}\right|^{-1}
$ and ${\bf \Gamma}=\Delta {\bf \Omega}$. The Fourier transform
of the kernel (\ref{eq:perker}) is therefore given by
\begin{equation}
\tilde K_{\bf I}({\bf \Phi};t)=
\sum_{\bf k} \tilde p_{\bf I} (t {\bf \Phi}) \exp(i {\bf \Phi}
({\bf \Gamma} t - 2 \pi {\bf k})).
\end{equation}
Then the formula
\begin{equation}
\sum_{\bf k} \exp(-i 2\pi {\bf \Phi}{\bf k})=\sum_{\bf j} \delta({\bf \Phi}-{\bf j})
\end{equation}
leads to
\begin{equation}
\tilde K_{\bf I}({\bf \Phi};t)=
\sum_{\bf j} \tilde p_{\bf I} (t {\bf j})\ \exp(it 
{\bf \Gamma}{\bf j} )\ \delta({\bf \Phi}-{\bf j}).
\end{equation}
This result, coupled with equations (\ref{eq:fidfour}) and (\ref{eq:evolfour}),
finally leads to
\begin{equation}
f(t)=\sum_{\bf j} |\tilde \rho_0({\bf j})|^2\ 
\tilde p_{\bf I}(t {\bf j})\ 
\exp(it {\bf \Gamma}{\bf j} ).
\end{equation}
As we can see, the behavior of fidelity in the limit   
$t\to \infty$ is given by the tails of the
Fourier transform of the kernel $p_{\bf I}$. The origin of the kernel 
is the projection of the perturbed tori onto unperturbed ones, 
and we expect singularities in such a projection. These singularities
induce a power law decay in the tails of the Fourier transform
of the kernel and thus are responsible
for the asymptotic power law decay of fidelity. 

In the single degree of freedom situation, the typical singularity 
of projection to be encountered leads to 
$p_I(y)\propto |y-y_0|^{-1/2}$, 
as it can also be seen in the twist map example (\ref{eq:distskew}). 
This type of singularity leads to the Fourier transform
\begin{equation}
\tilde p_I(\Phi)\propto \Phi^{-1/2} \exp(-i \Phi y_0).
\end{equation}
Such an expression leads to following asymptotic fidelity decay: 
$$
f(t)-f(\infty)=
$$
\begin{equation}
t^{-1/2} \sum_{j \neq 0} |\tilde \rho_0(j)|^2\ 
j^{-1/2} \exp(i j (\Gamma t - y_0))=
t^{-1/2} z(\beta),
\label{asymptotic}
\end{equation}
where $\beta=-\Gamma t + y_0$ and $z$ is some periodic function 
with period $2 \pi$.
We note that Eq.~(\ref{asymptotic}) gives an overall 
$\propto t^{-1/2}$ fidelity decay together with a 
superimposed oscillatory behavior.
This is the typical asymptotic relaxation of fidelity for 
a single torus in integrable systems with a single degree of freedom.
If one considers a finite interval of actions $\nu_I$, 
the decay (\ref{asymptotic}) must be averaged over $\nu_I$,
and therefore, due to the oscillatory nature of
Eq.~(\ref{asymptotic}), it can be faster that $t^{-1/2}$.
The extension to the many-dimensional case requires a complex   
analysis of the singularities encountered in the projection 
of the perturbed tori onto the unperturbed ones and is 
beyond the scope of the present paper.



\end{multicols}

\end{document}